\documentclass[aps,twocolumn]{revtex4}

\usepackage{graphicx}
\usepackage{color}

\begin{document}

\bibliographystyle{apsrev}

\title{A Bayesian Approach to the Detection Problem in Gravitational Wave Astronomy}
\author{\surname {Tyson} B. Littenberg and {Neil} J. Cornish}

\affiliation{Department of Physics, 
  Montana State University, Bozeman, MT 59717}
\date{\today}

\begin{abstract}
The analysis of data from gravitational wave detectors can be divided into three phases:  search,
characterization, and evaluation. The evaluation of the detection - determining whether a candidate
event is astrophysical in origin or some artifact created by instrument noise - is a crucial step in
the analysis. The on-going analyses of data from ground based detectors employ a frequentist
approach to the detection problem. A detection statistic is chosen, for which background levels
and detection efficiencies are estimated from Monte Carlo studies. This approach frames
the detection problem in terms of an infinite collection of trials, with the actual measurement
corresponding to some realization of this hypothetical set. Here we explore an alternative,
Bayesian approach to the detection problem, that considers prior information and the actual data in hand. Our 
particular focus is on the computational techniques used to implement the Bayesian analysis.
We find that the Parallel Tempered Markov Chain Monte Carlo (PTMCMC) algorithm is able to address all
three phases of the anaylsis in a coherent framework. The signals are found by locating the posterior modes,
the model parameters are characterized by mapping out the joint posterior distribution, and finally,
the model evidence is computed by thermodynamic integration. As a demonstration, we consider the detection
problem of selecting between models describing the data as instrument noise, or instrument noise plus the
signal from a single compact galactic binary. The evidence ratios, or Bayes factors, computed by the PTMCMC
algorithm are found to be in close agreement with those computed using a Reversible Jump Markov Chain Monte
Carlo algorithm.
\end{abstract}

\maketitle

\section{Introduction}
The first direct detection of gravitational waves (GWs) is now within reach, harkening the beginning of a new field of astronomy. 
The \emph{Laser Interferometer Gravitational wave Observatory} (LIGO) \cite{LIGO1, LIGO2} has recently completed acquiring one year
of triple-coincidence data at design sensitivities (S5) and analysis efforts are in full swing~\cite{S5}.  

Upgrades to the LIGO and Virgo \cite{Virgo} detectors are nearing completion in preparation for the S6/VSR2 data collection period,
paving the way for the advanced generation of observatories and routine gravitational wave detection. Plans for the space
based \emph{Laser Interferometer Space Antenna} (LISA)~\cite{LISA} continue to mature. All key LISA technologies are now in place
and will soon be flown on the \emph{LISA Pathfinder} \cite{LISAp} mission. While hardware developments have brought the
first detection within reach, exploting the full potential of these instruments requires similar improvements in the techniques
used to analyse the data. 

The ability to determine whether data contains just instrument noise, or noise plus a resolvable gravitational wave signal, is
the central problem in gravitational wave astronomy. The ``detection problem'' can be divided into three stages:

\begin{itemize}
\item \emph{Search}:  Determine the region(s) in parameter space containing the maximum posterior weight to establish a
candidate gravitational wave signal.
\item \emph{Characterization}: Establish confidence intervals for the parameters that describe the signal model and the noise model.
\item \emph{Evaluation}: The detection hypothesis is compared to alternatives through model selection.
\end{itemize}

Here we study a fully Bayesian approach to the detection problem.  All three stages of the analysis are handled by
a single technique - the Parallel Tempered Markov Chain Monte Carlo (PTMCMC) algorithm.  The PTMCMC algorithm
establishes which regions of parameter space contain the highest posterior weight, efficiently explores the posterior distribution
function (PDF) of the model parameters, and calculates the marginalized likelihood, or evidence, for the model. This
procedure is repeated for the competing hypotheses under consideration (e.g. instrument noise, instrument noise plus a
black hole inspiral signal, instrument noise plus $N$ galactic binary signals) and the evidence ratios, or Bayes factors,
are then used to identify the hypothesis that is most consistent with our prior belief and the current data. The Bayes
factors computed from the PTMCMC algorithm are checked against those computed using a Reverse Jump Markov Chain Monte Carlo (RJMCMC)
algorithm, and are found to be in very good agreement.

Several solutions to components of the GW detection problem have recently been successfully demonstrated using methods similar
to our own.  Parallel tempering has been shown to be effective in the characterization phase of ground-based data when
applied to simulated signals from spinning black hole binary inspirals~\cite{vanderSluys}.  A model selection study of when
an non-spinning inspiral waveform injected into simulated LIGO data has a sufficient signal-to-noise ratio to be favored over a
model containing only instrument noise has been performed using the Nested Sampling algorithm~\cite{Veitch1, Veitch2}. 

The paper is organized as follows:  In \S\ref{detection} we describe the detection problem and how it manifests differently
for ground- and space-based interferometers.  In \S\ref{bayes} we briefly describe the Bayesian approach to gravitational wave
data analysis and show how it can be used to characterize a candidate detection (parameter estimation) and to evaluate the
detection confidence (model selection).  In \S\ref{problem} we set up an example problem of a single galactic binary signal injected
into simulated LISA data and describe the parameterization of the models we wish to distinguish.  In \S\ref{implementation} we
explain the details of the search and characterization phase of the analysis.  We briefly introduce MCMC techniques followed
by an in-depth description of our specific implementation, including the prior and proposal distributions used in the MCMC algorithm
and the use of parallel tempering to thoroughly sample from the posterior distribution function for the model in question
(\S\ref{mcmc}) allowing us to simultaneously complete the search and characterization.  The evaluation of the candidate detection follows in
\S\ref{evaluation} where we show how to select between the competing models via thermodynamic integration (\S\ref{TI})
across the PTMCMC chains.  In \S\ref{model selection} we introduce an additional method for calculating the evidence ratio, the RJMCMC algorithm (\S\ref{rjmcmc}).  In \S\ref{results} we present the results of
our study, including a demonstration of the search efficiency, a determination of the signal-to-noise ratio at which
galactic binaries become distinguishable from instrument noise, and consistency checks applied to the computation of the
Bayes factors.  In \S\ref{discussion} we discuss the challenges associated with implementing these techniques, as well as the
innovations which allowed us to achieve the desired reliability.  We then outline the simplifications that went into this work
and what impact they will have when included in more general detection problems.

\section{The Detection Problem}  \label{detection}

\subsection{Ground Based Detectors}
For ground based GW detectors the detection problem exists because strong GW signals are rare, while large-amplitude transient noise
events in the detectors are common.  One must then be able to distinguish between these noise ``glitches" and GWs.  The majority
of glitches can be excluded by looking at auxiliary instrument monitoring channels (data quality tests and vetoes in LIGO parlance),
and by applying coincidence tests among the world-wide network of detectors. 

As early as 1989 it was suggested by Davis~\cite{Davis} that the gravitational wave detection problem may best be addressed
in a Bayesian framework. Indeed, many of the seminal papers~\cite{sam1,sam2,enna1,enna2,enna3,Anderson}
on gravitational wave data analysis employ Bayesian probability theory. These analysis are generally of a hybrid type, where Bayesian
reasoning is used to motivate the form of a particular frequentist statistic. This hybrid style of analysis is particularly evident in
recent studies of stochastic backgrounds~\cite{Allen}, unmodeled bursts~\cite{Searle}, methods for setting upper
limits~\cite{Biswas}, and targeted searches for continous signals~\cite{badri}. Standard choices of detection statistic
include signal-to-noise or likelihood ratios, and measures of correlation
across the network. Candidate events are identified using these statistics, and the significance of a given event is established
from Monte Carlo studies of signal injections and time slides of the data. The hypotheses that are being compared, and the
assumptions that are being made, are not always clearly spelled out in this approach, and in some cases may even be
unphysical~\cite{Searle, badri}.

There are several reasons why the Bayesian approach was not immediately adopted in practice, chief among them being the
computational challenge of evaluating the high dimensional integrals that appear in the analysis, and the lack of a reliable model
for the instrument noise. In recent years the development of efficient computational techniques such as Markov Chain Monte
Carlo methods (see e.g. Ref.~\cite{MCMC}) and Nested Sampling~\cite{Skilling1}, coupled with the widespread availability of cheap,
high speed compuational resources, have helped to address the computational challenge, but the problem of modeling the instrument
noise remains.

The Bayesian approach to model selection forces us to defined explicit models for the hypotheses under consideration. Once the
models are defined the analysis is entirely mechanical. The end result is an odds ratio for one hypothesis over another. 
We will see the Bayesian approach does not yield a fixed signal-to-noise threshold at which a particular class of signal becomes
detectable, but rather, the signal-to-noise ratio at which a particular instance of the signal in a given noise realization becomes
detectable can vary substantially. A note of caution here is the analysis is {\em not} answering the question ``is there a
gravitational wave signal present in the data, or is it just instrument noise?'', but rather, ``is the data most consistent with
our model of the the gravitational wave signal and instrument noise, or our model of the instrument noise alone?''. In particular,
if our model of the instrument noise is poor - such as not allowing for occasional glitches - then the odds may favor the detection
hypothesis even when no signal is present in the data. The goal is to find models for the signals and instrument noise~\cite{Allen}
that are sufficiently realistic so that the hypotheses that are being tested closely approximate the physical situations we hope
to compare. Here we focus on the development of a particulalry powerful technique - Parallel Tempered Markov Chain Monte Carlo -
to address the computational challenge of applying Bayesian Inference to gravitational wave data analysis, and defer the
development of more realistic noise models to a forthcoming publication.

\subsection{Space Based Detectors}
Whereas LIGO grapples with finding rare signals in a sea of noise transients, the space-based detection problem offers a different
set of challenges.  The most readily available sources expected to exist in the LISA bandwidth are the millions of binary star
systems comprised of stellar mass compact objects (mostly white dwarfs) within the galaxy~\cite{Nelemens}.  Hundreds of thousands of
such sources will have sufficient amplitude to measurably excite the LISA detector~\cite{Barak, Crowder2004}.  Binaries which have
already been detected electromagneticly can be used as verification sources for LISA by which the performance of the mission
can be evaluated. It should be possible to isolate tens of thousands of galactic binary signals from the LISA data, with
the remainder blending together to create a confusion background which is the dominant source of noise between $10^{-4}$ Hz and
$3 \times 10^{-3}$ Hz~\cite{Timpano}.  It is therefore critical to mission success that the maximum number of these sources can be
cleanly regressed from the data to allow access to the more exotic discoveries which will otherwise be obscured by the galactic
foreground.  

Many methods for attacking this problem have been developed and demonstrated in increasingly more sophisticated tests supplied by
the Mock LISA Data Challenge Task Force~\cite{MLDC}.  Because of the degree of overlap between the tens of thousands of resolvable
binaries, a global fit is needed~\cite{Cornish2005, Crowder2007}. Since the number of resolvable sources is {\em a priori} unknown,
the analysis must be able to find the best fit model along with the best fit parameters for a given model. Models with more
parameters will generally have higher likelihoods, but not necessarily higher evidence, as their prior distributions
are spread over a larger volume. Thus, a Bayesian approach to LISA data analysis automatically selects the most parsimonious model.

\section{Bayesian Inference}  \label{bayes}
Bayesian probability theory is slowly making its way into the mainstream of astrophysics and astronomy \cite{Gregory,Liddle, Trotta, Trotta2}.
In the Bayesian framework, the data $d$ is used to update our prior belief $p(\mathcal{H})$ in hypothesis $\mathcal{H}$ according to
the (marginalized) likelihood $p(d|\mathcal{H})$ that our hypothesis could have generated $d$ under the performed experiment.
The posterior belief in the hypothesis is given by Bayes' theorem:
\begin{equation} \label{Bayes' theorem}
p(\mathcal{H}|d)=\frac{p(d|\mathcal{H})p(\mathcal{H})}{p(d)}
\end{equation}
For now $p(d)$ is an unimportant normalization constant.  Unfamiliarity with Bayesian techniques can often result in discomfort with the 
way prior belief in the hypothesis  under scrutiny impacts the result of the analysis.  This is actually a feature of Bayesian analysis as it
requires the user to disclose the assumptions made about the model being tested.  Searle \emph{et al} \cite{Searle} give an interesting discourse on the hidden assumptions made by some commonly used GW detection statistics.

In model selection applications, the hypotheses are the competing
models $\mathcal{M}$ and the posterior $p(\mathcal{M}|d)$ is the probability (density) that $\mathcal{M}$ is the appropriate
description of the data.  In practice this requires prior knowledge of \emph{all possible models} testable by our experiment,
which is impractical.  Instead we must ask our data more carefully constructed questions, such as ``how do two specific models
compare?"  Since the normalization in eq. \ref{Bayes' theorem} is the same for all models ratios of the posterior odds eliminate
the need to know all hypotheses \emph{a priori} and yields a useful quantitative measure of confidence, the odds ratio, for one model
over another: 
\begin{equation}
\mathcal{O}_{10} \equiv \frac{p(\mathcal{M}_1|d)}{p(\mathcal{M}_0|d)}=\frac{p(\mathcal{M}_1)p(d|\mathcal{M}_1)}{p(\mathcal{M}_0)p(d|\mathcal{M}_0)}
\end{equation}
$\mathcal{O}_{10}$ is the odds ratio for models $\mathcal{M}_1$ and $\mathcal{M}_0$.  An odds ratio of $\gtrsim 3$ 
favors $\mathcal{M}_1$ \cite{Raftery}.  Evaluating the Bayes factor (also known as the evidence- or marginal likelihood-ratio) 
\begin{equation} 
B_{10} \equiv \frac{p(d|\mathcal{M}_1)}{p(d|\mathcal{M}_0)}
\end{equation}
is the goal of the evaluation phase of the analysis.  $B_{10}$ is interpreted as the degree of confidence in the appropriateness
of one model versus another in describing the available data.  The prior odds
$\mathcal{P}_{10} \equiv p(\mathcal{M}_1)/p(\mathcal{M}_0)$ reflect any preference between models before the experiment is conducted.
If we are to adopt one model over the other, the Bayes factor must overwhelm the prior odds, thus signaling the data is sufficiently
informative to distinguish between the competing hypotheses.

For the current generation LIGO detection problem $\mathcal{P}_{10}$ would likely down-weight the chances of a resolvable GW signal
based on some combination of reasonable event rates and detector behavior around the time of the candidate detection.  This is
analogous to setting a relatively high threshold which must be exceeded for a candidate detection to warrant further scrutiny.  For the
LISA extension of this same problem (due to the large number of anticipated detections) the prior odds would likely be far less
restrictive, leaving the analysis more open to the possibility of a signal being present.

Advocates of Bayesian inference cite a natural parsimony in the model selection application akin to Occam's Razor.  The claim is
that if two models give similar quality ``fits'' to the data the lower dimensional model will be preferred.  This line of reasoning
is only valid when accompanied by a strict qualifying statement:  \emph{Excess model parameters must be constrained by the data}.
The marginalized posterior distribution function of an unconstrained parameter will match the prior distribution, meaning the data
was incapable of updating our understanding of that parameter and can not be used to answer questions about whether or not that
quantity ``belongs" in the model (See Ref.~\cite{Liddle1} for an excellent discussion of this and other misconceptions regarding Bayesian inference).  That model then incurs no penalty for having unconstrained variables included in its
parameterization. If the remaining parameters do as well describing the data as some other, lower dimensional model, both are
equally viable.  Reverend Bayes brandishes his razor only if a particular model \emph{requires} more parameters to achieve a
similar fit.  

The ``penalty'' in the Bayes factor is roughly equal to the ratio between the widths of the posterior and the prior distributions
of the additional parameters.  If this is near unity than there is no penalty for the additional dimensions and the two models are
indistinguishable.  Tightly constrained model parameters will incur a large Occam factor and thus must substantially improve the
likelihood in order for that model to be preferred.  This is a somewhat subtle and often overlooked point that, when addressed,
enhances the conceptual understanding of Bayesian model selection.

For parameter estimation applications, the hypothesis states that the acquired data is the result of some process parameterized
by $\vec{\theta}$.  The posterior distribution function $p(\vec{\theta}|d, \mathcal{M})$ is the probability density function
for the model parameters over the parameter space.  Producing a well sampled PDF is the goal of the search and characterization
phase of the analysis. The global maximum of the PDF, or maximum a posteriori (MAP) point, occurs where the model parameters are the
``best fit'' values for the hypothesis in question (weighted by the prior belief in those parameter values).  The distribution
about this location in parameter space reveals how well the data constrains the hypothesis. Using Bayes' theorem we have
\begin{equation} \label{Bayes2}
p(\vec{\theta}|d, \mathcal{M})=\frac{p(d|\vec{\theta}, \mathcal{M})p(\vec{\theta}, \mathcal{M})}{p(d | \mathcal{M})}
\end{equation}
In this instance $p(\vec{\theta}, \mathcal{M})$ is the prior distribution of the model parameters, and $p(d|\vec{\theta}, \mathcal{M})$
is the likelihood that the model could have generated data $d$ for model parameters $\vec{\theta}$. The normalization constant
\begin{equation} \label{evidence integral}
p(d|\mathcal{M}) = \int d\vec{\theta} \ p(\vec{\theta},\mathcal{M})p(d|\vec{\theta},\mathcal{M})
\end{equation}
is the marginalized likelihood, or evidence, for the model. In practice it quickly becomes too costly to perform a direct
evaluation of the posterior distribution function. In typical GW data analysis applications the number of parameters needed
to describe the signal model and noise model can be very large (ranging from tens to hundreds of thousands), and the regions of
significant posterior weight generally occupy a minute fraction of the total prior volume. Resolving the peaks in the
posterior distribution function requires a very fine sampling of the parameter space, but if this sampling is extended over
the entire prior volume the total number of samples can become astronomically large.

The Markov Chain Monte Carlo approach encompasses a powerful set of techniques for producing samples from the posterior
distribution. These algorithms focus their attention on regions of high posterior weight, and neatly avoid the problem
of computing model evidence. The later feature becomes less desirable when it comes time to test competing hypotheses.
One solution is to generalize the transitions between steps in the Markov Chain to include ``trans-dimensional'' moves
between different models, resulting in what are termed Reverse Jump Markov Chain Monte Carlo (RJMCMC) algorithms. The number of
iterations that the chain spends in models $\mathcal{M}_1$ and $\mathcal{M}_0$ provides an estimate for the evidence ratio
$B_{10}$. A more complete description of this approach can be found in \S\ref{rjmcmc}.

In what follows we will focus on a comprehensive approach to Bayesian data analysis that combines the Parallel Tempered Markov Chain 
Monte Carlo algorithm for conducting the search and performing the characterization with thermodynamic integration across the chains 
providing a robust estimate for the model evidence.

\section{Toy Problem}
\label{problem}
To test our analysis scheme we want to use a toy problem that allows for fast computation of likelihoods, while still
being fairly realistic. To that end we generate a 256 frequency-bin segment of simulated LISA data which contains
instrument noise and the signal from a single chirping galactic binary over an observation time of $T_{\rm obs} = 2$ years.   

Ideal GW data $s$ has two independent contributions:  The incident gravitational radiation convolved with the transfer
function of the detector generates some response $h$ in addition to noise $n$ caused by some collection of random processes
within the instrument. This can be written in the Fourier domain as
\begin{equation}
\tilde{s}_{\alpha}(f) = \tilde{h}_{\alpha}(f) + \tilde{n}_{\alpha}(f)
\end{equation}
where the subscript ${\alpha}$ indicates the output specific to a single detector (in the case of the ground-based effort) or
a particular interferometer channel of a single detector (as for LISA).  The LISA detector will be an array of three satellites,
each housing a pair of test masses and the appropriate optical system to precisely monitor the distance between itself, the test
masses, and the other two satellites.  The relative distance measures between each satellite will be combined to form three
coupled Michelson-type interferometer channels.  We will work in the noise-orthogonal time-delay interferometry basis which is
formed out of linear combinations of the Michelson channels, commonly referred to as channels A, E, and T \cite{AET}.
This choice has the added benefit of reducing to only two data channels of interest, as T is free of GW signals within the
expected frequency range of galactic binaries.  

We first generate a noise realization for the frequency range of interest by drawing from the theoretical LISA noise
spectral density (see \S\ref{noise model}).  We then inject a test source somewhere in the middle third of the data
with some tunable amplitude so that we may establish at what signal to noise ratio (SNR) the source becomes detectable.
To be considered effective our algorithm must find the MAP parameters for two distinct models 
\begin{eqnarray}
\mathcal{M}_0:  \tilde{s}(f) &=& \tilde{n}(f)  \nonumber \\
\mathcal{M}_1:  \tilde{s}(f) &=& \tilde{n}(f) + \tilde{h}(f), 
\end{eqnarray}
thoroughly resolve each models PDF, and calculate the evidence for each model (up to some constant factor common
between $\mathcal{M}_1$ and $\mathcal{M}_0$).  We will consider the prior odds between the two models to be unity,
thus allowing us to interpret the evidence ratio (Bayes factor) as the odds ratio.

\subsection{Modeling the Source:  Galactic Compact Binaries}
The choice of using galactic binaries as the test source was motivated by the fact that waveforms for these objects
can be rapidly modeled using the fast-slow decomposition described in \cite{Tyson}.  Despite the simplicity of generating
these waveforms, they posses sufficient complexity that we do not anticipate significant additional complications related to the
modeling/fitting of other, more exotic sources (the same techniques described here, with a few enhancements, are proving
sucessful at handling spinning black holes and extreme mass ratio inspirals).

Detached, mildly chirping compact binaries are well modeled, and can be parameterized using eight quantities which we will
consider as components of a parameter vector $\vec{\lambda}$:
\begin{equation}
\vec{\lambda} \rightarrow (\ln A, fT_{\text{obs}}, \dot{f}T_{\text{obs}}^2, \cos \theta, \phi, \psi, \cos \iota, \varphi_0)
\end{equation}
The physical parameters are the amplitude $A$, GW frequency $f$ (twice the binary's orbital frequency), frequency derivative
$\dot{f} =df/dt$, co-latitude and azimuthal sky angles $\theta$ and $\phi$, GW polarization angle $\psi$, orbital
inclination $\iota$, and phase $\varphi_0$ of the gravitational waves.  The quantities $f$, $\dot{f}$ and $\varphi_0$ are
defined at some fiducial time such as when the first data is taken.

\subsection{Modeling the Instrument Noise} \label{noise model}
The contribution to the data from instrument noise in each Fourier bin, $\tilde{n}(f)$, has spectral density $S_n(f)$
with mean and variance
\begin{eqnarray}
\langle \tilde{n}(f)_i \rangle &=& 0 \nonumber \\
\langle \tilde{n}(f)_i \tilde{n}(f)_j \rangle &=& \frac{T_{\rm obs}}{2}\delta_{ij} S_n(f)
\end{eqnarray}
respectively.  For operational detectors this is approximately true but excursions from the expected level are to be anticipated.
In preparation for this reality during the LISA mission we developed our algorithm to simultaneously fit for the noise level in the
data.  We parameterize the noise by assuming that departures from $S_n(f)$ can be characterized by a rescaling of the noise
spectral density.  This rescaling is unique to each channel (A and E) while $S_n(f)$ is identical for both.  A more sophisticated
approach would be to fit for the twenty-four components of noise in the LISA constellation (the laser phase noise between, and
the position noise of, each test mass).  This approach has been developed by Adams and Cornish~\cite{Adams}, but has not
been implemented here.  

\begin{figure}[htbp]
   \centering
   \includegraphics[width=0.5\textwidth]{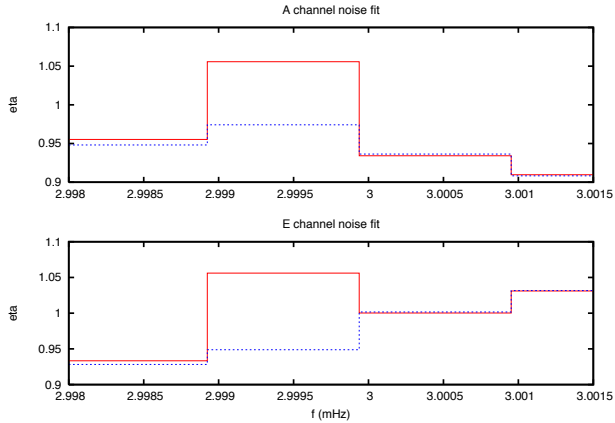} 
   \caption{MAP noise parameters for $\mathcal{M}_0$ (red, solid) and $\mathcal{M}_1$ (blue, dashed).  The data consists of two interferometer channels, each containing 256 Fourier bins which are divided into four sub-regions.  Each sub-division is fitted with a unique noise parameter.  The signal is injected somewhere in the second quarter of the data.  Model $\mathcal{M}_0$ elevates the noise parameter in the second window to account for the excess power caused by the gravitational wave signal.  Model $\mathcal{M}_1$ successfully fits to that gravitational wave leaving the noise parameter closer to unity.  Notice how the noise parameters for remaining portions of the data are nearly identical between the two models.}
   \label{noise plot}
\end{figure}

The simulated segment of LISA data is divided into four equal length sub-regions in frequency space of size $\Delta f$. 
Since the noise spectral density is reasonably constant over the entire data segment the central frequency is used to evaluate
$S_n(f)$ which will furthermore be treated as constant, while the noise level in each sub-region $i$ and interferometer
channel $\alpha$ is rescaled by $\eta^i_{\alpha}$ to yield the modeled noise level
\begin{equation}
S_n^{\alpha}(f) \rightarrow \eta_i^{\alpha}S_n.
\end{equation}
For the case where there are two non-negligable interferometer channels (A and E) a total of eight noise parameters
are required.  An example of these parameters for a particular noise realization can be seen in Fig. \ref{noise plot}.  

We have now fully parameterized two competing models 
\begin{eqnarray}
\mathcal{M}_0: \vec{\theta}_0 &=& \sum_j\vec{\eta}_j  \nonumber \\
\mathcal{M}_1: \vec{\theta}_1 &=&  \vec{\lambda} +  \sum_j\vec{\eta}_j 
\end{eqnarray}
which can be used to explain the data in question.  Our attention will subsequently turn to applying Bayesian tools to
establish the most appropriate configuration of each model given the simulated data, and to learn which model most
appropriately describes the data in question.

\section{Implementation of Bayesian Techniques:  Detection \& Characterization} \label{implementation}
\subsection{Markov Chain Monte Carlo} \label{mcmc}
Markov Chain Monte Carlo (MCMC) \cite{Metropolis, Hastings, MCMC} techniques are becoming quite familiar in the
GW community~\cite{nelson} thanks to their great effectiveness in solving parameter estimation problems.
Variants of these techniques have also proven
quite effective as search algorithms~\cite{Cornish2005, Crowder2007, Cornish2007, Cornish2008, Gair}.  Here a brief description of
the MCMC technique is followed by a more in-depth discussion of the particular implementation that we used.

The utility of the MCMC approach stems from its ability to explore the parameter space in such a way that the resulting chain
samples directly from the target posterior distribution function.  This is accomplished by first adopting some position in
parameter space $\vec{x}$ as the first ``link" in the chain.  A subsequent move to a new position $\vec{y}$  is proposed using
some (normalized) distribution $q(\vec{y}|\vec{x})$ read \emph{``the probability of moving to $\vec{y}$ given that the current
location is $\vec{x}$"}.  The new likelihood $p(s|\vec{y})$ and prior probability $p(\vec{y})$ are evaluated and the new position
is accepted with some probability ${\kappa} =\text{min}[1,H]$ where $H$ is the Hastings ratio
\begin{equation}
H_{\vec{x} \rightarrow \vec{y}}=\frac{ p(s|\vec{y}) p(\vec{y}) q(\vec{x}|\vec{y}) }{ p(s|\vec{x}) p(\vec{x}) q(\vec{y}|\vec{x}) }.
\end{equation}
The Hastings ratio is designed to ensure detailed balance, and hence reversibility, between subsequent steps in the chain.
This ensures free exploration of the PDF without any bias or hysteresis.  The process of stochastically stepping through
parameter space is repeated until some convergence criteria is satisfied.  Afterwards, the number of iterations spent in a
particular region of parameter space, normalized by the total number of steps in the chain, yields the probability that the
source parameters take values in that region.

Marginalized posterior distribution functions can then be inferred by constructing histograms of the parameter values (discarding
early iterations from the ``burn in" phase before the chain reaches stationarity).  A common practice is to choose a proposal
distribution that is symmetric, thus eliminating the need to include it in the calculation of $H$.  The choice of
$q(\vec{y}|\vec{x})$, by construction, \emph{can not} alter the recovered posterior distribution function. The proposal distribution
does, however, dramatically affect the acceptance rate of trial locations in parameter space and, therefore, the number of
iterations required to satisfactorily sample the joint PDF.

\subsubsection{Evaluating the Likelihood} \label{likelihood}
The standard likelihood definition used in matched filtering searches with ideal (stationary and Gaussian) noise is
\begin{equation} \label{likelihood definition}
p(s|\vec{\lambda}) \equiv C e^{-\frac{1}{2}\chi^2}
\end{equation}
where $\chi^2 = (s-h(\vec{\lambda})|s-h(\vec{\lambda}))$ and the normalization constant $C$ is independent of the source
parameters~\cite{Finn}. It does however depend on the noise parameters. The brackets $(\cdotp|\cdotp)$ denote a noise weighted
inner product which, for a finite observation
time $T_{\rm obs}$, is defined as
\begin{equation} \label{nwip}
(a|b) \equiv \frac{2}{T_{obs}}\sum_{\alpha} \sum_f \frac{ \tilde{a}^*_{\alpha}(f)\tilde{b}_{\alpha}(f) 
+ \tilde{a}_{\alpha}(f)\tilde{b}^*_{\alpha}(f) }{ S_n^{\alpha}(f) }
\end{equation}
with the sum over $\alpha$ indicating the presence of multiple interferometer channels contributing to the data.
These can be from different detectors (\emph{e.g.} LIGO H1, H2, and L1) or different combinations of channels from
a single interferometer (LISA).

In our implementation the noise spectral density is not fixed, and care must be taken to account for this when
computing the likelihood. In particular, the normalization of the likelihood depends on the noise parameters. By elevating
the non-constant components of $C$ into the argument of the exponential the likelihood is now written as
\begin{equation}
p(s|\vec{\lambda}, \vec{\eta}) \equiv C'\exp \Big{[} -\frac{1}{2}\Big{(}\chi^2 + N\sum_{\alpha,j} \ln \eta^{\alpha}_j \Big{)} \Big{]}
\end{equation}
where $N=T_{\rm obs}\Delta f$ is the number of Fourier bins over range $\Delta f$ and the sum over $j$ ranges over the number
of $\Delta f$ segments in the data.  Meanwhile our definition of the noise weighted inner product (eq. \ref{nwip}) must also reflect
the parameterization of the noise by replacing $S_n^{\alpha}(f)$ with $\eta^{\alpha}_i S_n$.

\subsubsection{Prior Distributions}\label{prior}

One of the selling points for gravitational wave astronomy is that it will open a window on astrophysical sources where
relatively little is currently known. Consequently, the prior distributions for the parameters in our models will tend to
be fairly wide. For the problem at hand there are population synthesis models that combine our best understanding of the
physical processes that govern stellar evolution and galactic structure, but a paucity of observational data, especially
for short period systems, mean that there are considerable uncertainties. One could combine the available models
to produce predictions for the joint probability of systems having a certain period, sky location, distance from the Earth,
masses, and mass transfer rate, then translate these into a joint prior for $A,f,\dot{f},\theta$, and $\phi$. The orientation of
the orbit is expected to be random, as is the initial orbital phase, so the parameters $\psi, \cos \iota$, and $\varphi_0$ should
be uniform across their allowed range. These astrophysical priors may be supplemented by considerations of the observational
biases, such as the inability to detect sources below a certain amplitude threshold.

For our toy problem we simplified the analysis by adopting uniform priors for $\ln A, f, \cos\theta$, and $\phi$. We did however
use a particular population synthesis model to provide a prior for the frequency derivative, after marginalizing over the
other model parameters. For a detailed description of how this prior is constructed, and its effect on parameter estimation,
see Ref.~\cite{Tyson}. Save for the amplitude, all of the parameters in our model have a natural range, so it is easy to
construct properly normalized prior distributions. The choice of range for the amplitude requires more thought.

For the amplitude we used a very conservative lower limit for what might be detectable, and an upper limit that is
much larger than any of the injections used in the study. A rough estimate for the signal-to-noise ratio (SNR) of
a source with amplitude $A$ can be derived by assuming that all the signal power is deposited in a single Fourier
bin:
\begin{equation}
{\rm SNR}_{\rm est} = \frac{A \sqrt{T_{\rm obs}}}{\sqrt{S_m(f)}} \, ,
\end{equation}
where $S_m(f)$ is the noise spectral density for a Michelson-like interferometer channel, which is related to
the noise in the $A,E$ TDI channels by
\begin{equation}
S_m(f) = \frac{S_n(f) }{ 4 \sin^2(f/f_*) }  \, .
\end{equation}
Unless otherwise stated, we used a uniform prior range on $\ln A$ such that estimated signal-to-noise ratio lay in the
range ${\rm SNR}_{\rm est}\in [1, 20]$. The lower end of this range is certainly not detectable, so we expect to find
that the model selection procedure will be unable to strongly favor the noise model over the signal+noise model,
as the signal model extends to what is effectively zero amplitude. To explore the impact of the lower bound
we will also consider the effect of setting an amplitude threshold corresponding to ${\rm SNR}_{\rm est} = 5$.
Put another way, the non-detections of gravitational waves by ground based instruments does not rule out the
existence of gravitational waves, but rather, sets limits on the event rate and strength of GW signals.

It is expected that the results of the analysis will vary depending on the choice of prior distributions, but having
uninformative priors also yields a significantly more difficult problem to solve. The fraction of the prior volume
which contains the majority of the support for the posterior distribution function governs the difficulty of the analysis.
For galactic binaries, which have relatively small dimension and simple waveforms when compared to more exotic LISA sources
(such as extreme mass-ratio inspirals and spinning binary black holes) we find that, for marginally detectable signals
in our toy problem, $\sim 90\%$ of the posterior mass is contained within $\sim 10^{-15}$ of the prior volume.
That is what makes the search for gravitational wave signals so difficult. The ratio is many orders of magnitude smaller for
more complex signals.  An additional complication
is that the parameter volume that contains $\sim 90\%$ of the posterior mass is typically comprised of multiple disjoint
regions, and the analysis tools must be able to explore all of these posterior modes.

\subsubsection{Proposal Distributions}
The scheme for proposing trial positions in the parameter space has no bearing on the ultimate results of any MCMC analysis
but it greatly affects the amount of time required to arrive at that solution.  Convergence time is at a minimum if the
proposal distribution matches the posterior distribution.  Although we can not expect to know the ``answer'' (the PDF) until
after we've done the analysis, physical considerations can help in the design of proposal distributions that approximate the
posterior.  The noise-model parameters are independent of one another meaning each should have a PDF described by
a chi-squared distribution with $N$ degrees of freedom ($N$ is the number of samples in the data window where the noise parameter
is applied, see \S\ref{likelihood}).  For a sufficiently large number of frequency bins per window the chi-squared distribution
can be approximated as a normal distribution with variance $\sigma_{\eta}^2=1/N$ .  New noise parameters can then be proposed
by drawing from this distribution via:
\begin{equation}
\eta_y^i = \eta_x^i  + N[0,1]\frac{\sigma_{\eta}}{\sqrt{D_{\eta}}}
\end{equation}
where $N[0,1]$ is a normal distribution with zero mean and unit variance, $D_{\eta}$ is the noise-model dimension and is
used to rescale the jump to $\eta_y$.  This ensures the total noise parameter jump to be typically one standard deviation
from the mean of the presumed joint noise posterior distribution function.

The signal-model parameters are known to have non-zero correlation so efficient proposal distributions should be
constructed using the inverse covariance matrix of the parameter space.  This is commonly approximated by the
Fisher Information Matrix ($\bf{\Gamma}$) with components
\begin{equation}
\Gamma_{ij}=(h_{,i}|h_{,j}).
\end{equation}
Here  $h_{,i}$ denotes a partial derivative of the template $h(\vec{\lambda})$ with respect to the $i^{\text{th}}$
parameter \cite{Cutler}.  Jumps in parameter space are then made in the eigen-directions of the Fisher by drawing from
$N[0,1]$ and scaling it by the associated eigen-value.  These jumps are then additionally rescaled by $1/\sqrt{D_{\lambda}}$,
where $D_{\lambda}$ is the signal-model dimension to again achieve typical jumps of one standard deviation.   Instead of re-evaluating the matrix at every step of the chain, we keep the same Fisher for many iterations (making for a symmetric proposal distribution), under the assumption that it does not vary greatly in the vicinity of a local maximum.  Although this type of proposal distribution very efficiently samples the posterior in the vicinity of the signal
parameters (for sufficiently high SNR) it is a poor choice for sampling the entire posterior, as large jumps are
very unlikely to occur.  

Since the Fisher Information Matrix is a local approximation to the inverse covariance matrix, it can not be used to predict
the parameter correlations far from the current location.  For larger jumps we use a simpler proposal that takes
the estimated variance for each parameter (as determined from $\bf{\Gamma}$) and use this to scale additional draws
from $N[0,1]$ to jump in parameter directions.  This proposal distribution cocktail is then rounded out by occasionally
proposing uniform draws on the prior, ensuring that trial samples cover the entire prior volume during the exploration of
the target distribution. This collection of proposal distributions have the added benefit of being symmetric in $\vec{\theta}_x$
and $\vec{\theta}_y$, allowing us to neglect their contribution to the Hastings ratio.

\subsubsection{Parallel Tempering} \label{parallel tempering}

Although our proposal distributions permit exploration of the entire prior volume, the bolder jumps are rarely accepted as
the chance of making large changes to a large number of parameters and still ending up at a location with decent likelihood
is vanishingly small.  Furthermore, it is our desire that this Markovian chain can also be used to \emph{find} the injected
signal even if it is started randomly within the prior volume.  A straight forward implementation of the MCMC algorithm will
never (practically) find the MAP for parameter spaces of this complexity without some assistance.  Usually this assistance
comes in a way that violates the detailed balance condition between subsequent iterations and thus nullifies the Markovian
nature of the chain (resulting in biased samples which do not mirror the PDF).  This is of no negative consequence as long as
these non-Markovian steps in the chain are discarded as burn-in samples, but it does put tremendous pressure on the burn-in to
put the chain very near to the global maximum.

To encourage efficient global sampling of the target distribution, which in turn expedites the convergence time of the chain,
while preserving detailed balance we have adopted the powerful technique of parallel tempering~\cite{PTMCMC}, where multiple
chains explore the data simultaneously, each sampling from an iteratively higher ``temperature" target distribution 
\begin{equation}
p(\vec{\theta}|s,\beta) = p(\vec{\theta})p(s|\vec{\theta})^{\beta}
\end{equation}
where $\beta$ is analogous to an inverse temperature.  This effectively ``softens" the likelihood surface allowing the chain
easier access to positions in parameter space that would otherwise be difficult to reach.  Information from ``hot" chains is
allowed to propagate towards colder chains (and vice versa) by randomly proposing parameter exchanges between temperature levels.
An exchange of parameters between the $i^{\text{th}}$ and $j^{\text{th}}$ chain satisfies detailed balance if the conditional
probability is evaluated with Hastings ratio:
\begin{equation}
H_{i\leftrightarrow j}=\frac{p(s|\vec{\theta}_i, \mathcal{M}_i, \beta_{j})p(s|\vec{\theta}_{j},  \mathcal{M}_{j},\beta_{i})}
	    {p(s|\vec{\theta}_i, \mathcal{M}_i, \beta_{i})p(s|\vec{\theta}_{j}, \mathcal{M}_{j}, \beta_{j})}
\end{equation}
Only points in the $\beta=1$ chain sample the true posterior and are therefore permitted to contribute to the Markov chain
from which the PDF is inferred.  However, by exchanging parameters with the hot chains the $\beta=1$ chain rapidly explores
the full target distribution revealing a wealth of information about the posterior.  This is in contrast to chains without
parallel tempering which are prone to ``sticking'' at or near the maximum of the posterior, or worse, some secondary maxima.
Although non-parallel-tempered chains efficiently sample the region around some maxima they can completely miss
(for a finite run time) the more complex structure of the target distribution.  

\begin{figure*}[htbp]
\begin{center}
\includegraphics[width=\textwidth]{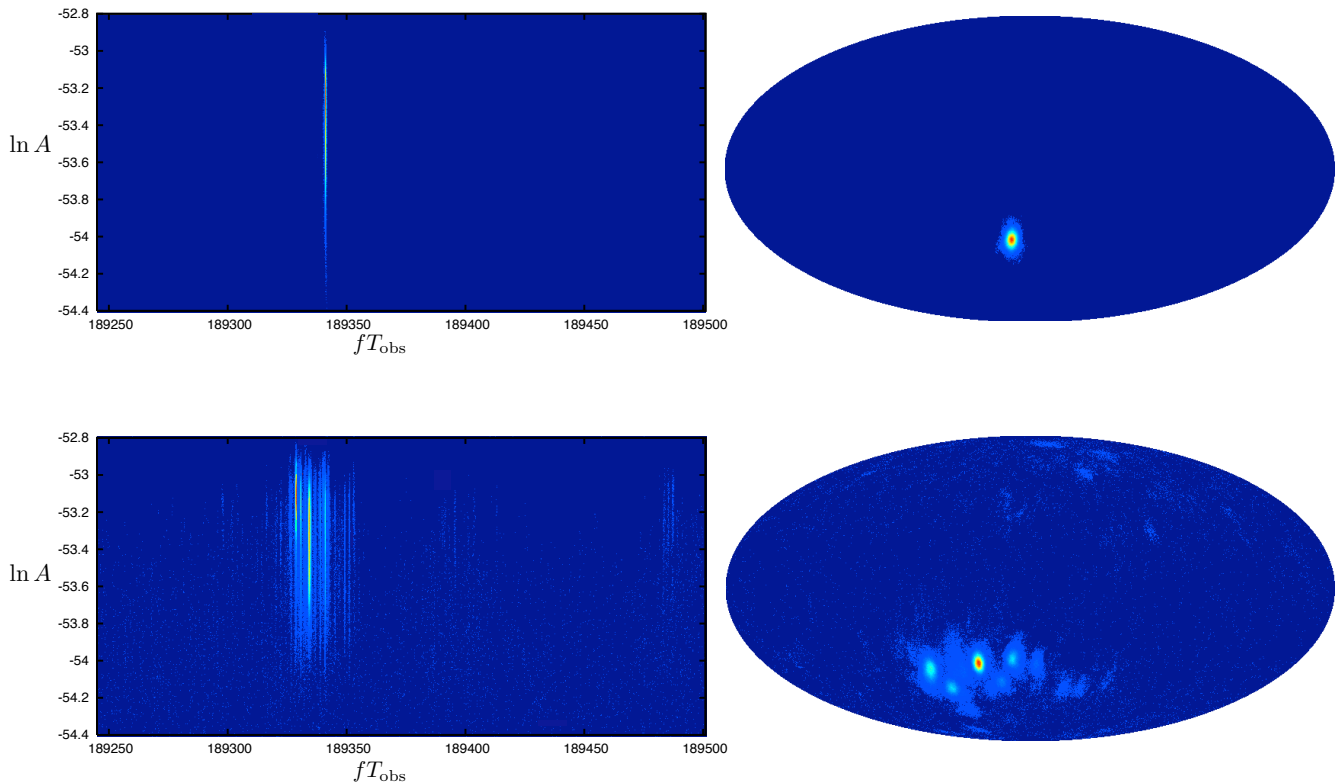}
\caption{Two dimensional marginalized log posterior densities for a single galactic binary with $SNR=7$ as approximated by a MCMC.  On the left is the $\ln A-fT_{\text{obs}}$ plane, the right shows a Molweide projection of the sky location (parameters $\theta$ and $\phi$). Red (white) locations have the highest log probability density while blue (black) has the least.  The top panel is for an MCMC without parallel tempering where the chain was started at the injected parameters and allowed $5\times10^5$ burn-in iterations.  The bottom panel is the same, only now with twenty parallel chains spaced geometrically in heat with a maximum temperature of 100 ($\beta_{\text{max}} = 0.01$).  The PTMCMC samples the full distribution, resolving the global structure of the PDF, while the straight-forward single chain MCMC never
leaves the region around injected parameter values, missing the global maximum entirely (which happens to be pushed off from the injected
values by instrument noise).}
\label{ptsky}
\end{center}
\end{figure*}

Although parallel tempering increases the computational cost of each iteration, for well
constructed heating schemes, $N_c$ chains running for $I$ iterations generically converges faster than one chain allowed to
run for $N_c \times I$ iterations.  Figure \ref{ptsky} clearly demonstrates the advantage of augmenting a typical MCMC with
parallel tempering.  The figure shows two dimensional marginalized posterior distribution functions as sampled by an MCMC as
described above.  The injected source was a single galactic binary with a SNR~$=7$.  For demonstrative purposes the chains
were initiated at the injected signal parameters and underwent $1.5\times 10^6$ iterations, the first $5\times10^5$ being discarded as burn in
samples.  The PDFs are for the $\ln A - fT_{\text{obs}} $ and  $\cos \theta-\phi$ planes.  The MCMC without parallel tempering
clearly samples the region around the injected parameters but is confined to that location in parameter space and would need to
run beyond a time that is practical in order to sample the full distribution.  The parallel tempered MCMC samples the full PDF
very efficiently, and for this example discovers a global maximum that significantly differs from the injected parameters
signaling how the MAP parameters can be pushed away from the injected parameters by the noise.  

This particular example also demonstrates the harm in being solely interested in the ``best fit'' parameters, as opposed
to the full distribution.  Had a search been performed over this data (as opposed to starting the chain at the injected parameters)
through some MCMC driven technique (such as using simulated annealing during the burn-in phase) without parallel tempering it is
likely that \emph{only} the global maximum will have been sampled by the post burn-in MCMC recovering signal parameters that are
measurably different from the physical parameters of the source which is producing the gravitational waves.  This may be of little
consequence for a single galactic binary, but for more exotic signals where source parameters will be used to learn about important
astrophysics, or where optical counterparts will be of great interest, it is clear that the full PDF is vital for complete
understanding of the underlying astrophysics.  Alternatively, a well constructed PTMCMC algorithm will simultaneously locate
maxima in the search space (satisfying the search phase of the detection problem) while accurately and efficiently mapping out
the posterior distribution function (characterizing the gravitational wave source).  

\section{Implementation of Bayesian Techniques:  Evaluation} \label{evaluation}
\subsection{Thermodynamic Integration} \label{TI}
Although only the $\beta=1$ chain samples the target distribution in the PTMCMC approach, the higher temperature chains serve
as more than just a convergence aid.   These additional chains can also be utilized to calculate the model evidence. 
For normalizable priors we can evaluate $p(d|\mathcal{M})$ from eq.~\ref{evidence integral} by using the expectation value
of the likelihood for each temperature level.  First we consider the evidence for each temperature's posterior distribution
function as part of some partition function $Z(\beta)$ where
\begin{eqnarray}
Z(\beta) &\equiv & \int d\vec{\theta}\ p(d|\vec{\theta},\mathcal{M},\beta) p(\vec{\theta},\mathcal{M}) \nonumber \\
&=&  \int d\vec{\theta}\ p(d|\vec{\theta,\mathcal{M}})^{\beta}  p(\vec{\theta},\mathcal{M}).
\end{eqnarray}
Because the prior is independent of $\beta$ we can write the partition function as
\begin{equation}
\frac{d}{d\beta} \ln Z(\beta) = \langle \ln p(d|\vec{\theta},\mathcal{M}) \rangle_{\beta}
\end{equation}
where $ \langle \ln p(d|\vec{\theta},\mathcal{M}) \rangle_{\beta}$ is the expectation value of the likelihood for the
chain with temperature $1/\beta$.  Then the evidence can be found by integrating over $\beta$ via
\begin{equation}
\ln p(d|\mathcal{M}) = \int_0^1 d\beta \ \langle \ln p(d|\vec{\theta},\mathcal{M})\rangle_{\beta}.
\end{equation}
Alternatively, after the data has been analyzed under both models the Bayes factor is calculated by
\begin{eqnarray}\label{thermo_bayes_integral}
\ln \mathcal{B}_{10} &=& \int_{-\infty}^0\beta( \langle \ln p(d|\vec{\theta}_1,\mathcal{M}_1)\rangle_{\beta} \nonumber \\
 &-& \langle \ln p(d|\vec{\theta}_0,\mathcal{M}_0)\rangle_{\beta}) \ d \ln{\beta}.
\end{eqnarray}
The expectation value of the likelihood is evaluated over the post-burn-in chain iterations~\cite{Gregory1, Gregory2}.
This technique, known as thermodynamic integration (TI), is a direct calculation (instead of an approximation ) to
the evidence~\cite{thermo}. It is not necessary to thermodynamic integration that the different temperature chains
be allowed to exchange parameters, however allowing them to do so ensures healthy mixing and convergence.

The evidence ratio between competing models $\mathcal{M}_1$ and $\mathcal{M}_0$ then yields the Bayes factor for the
models in question.  We now have a single algorithm that, if implemented effectively, solves all three facets of the
detection problem free of any approximations which may render the result suspect. 
\begin{figure}[htbp]
   \centering
   \includegraphics[height=0.5\textwidth, angle=270]{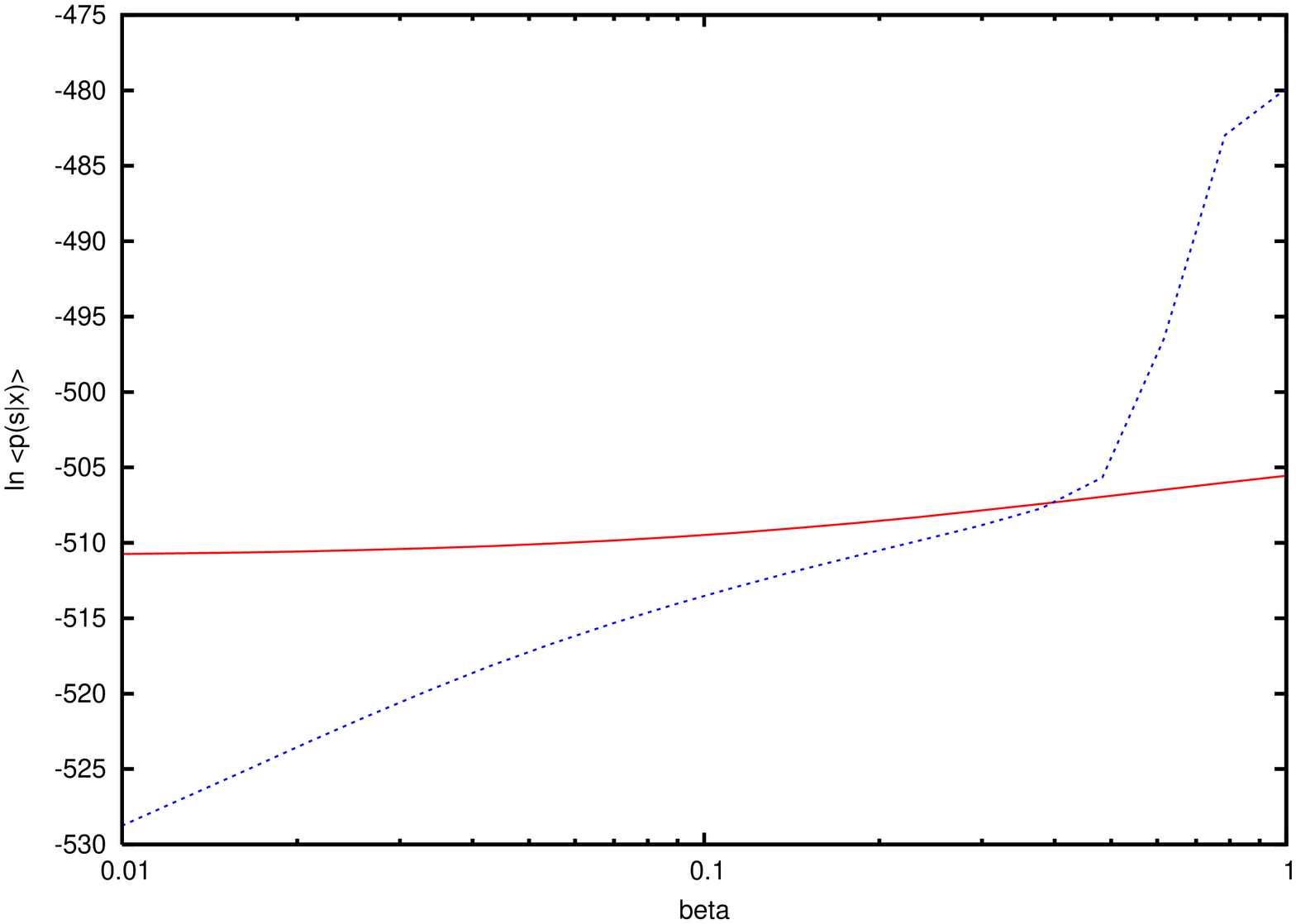} 
   \caption{Average likelihood for chains of different heats $\beta = 1/T$.  The red (solid) line is for the noise
only model $\mathcal{M}_0$ while the blue (dashed) line is for the signal plus noise model $\mathcal{M}_1$.
This particular example was for data containing a SNR $=\ 8$ source.  }
    \label{thermo}
\end{figure}

\section{Model Selection Cross Validation} \label{model selection}
\subsection{Reverse Jump Markov Chain Monte Carlo} \label{rjmcmc}
Although TI is robust and converges predictably, we feel that for ``high stakes'' model selection problems
(such as the first GW detection) several complimentary methods should be used to support the detection model
over any alternative hypotheses in order to establish additional confidence in the detection.  One approach for
such a case is to use alternative techniques for calculating the Bayes factor.  For this additional checking
stage we have chosen the ``gold standard'' of Bayes factor calculators:  The reverse jump Markov chain Monte Carlo
(RJMCMC) algorithm.  

This breed of MCMC was originally developed by P.J. Green in 1995 \cite{Green1995} and is unique in that it has
the ability to transition between competing models, effectively making the model one of the search parameters. 
This technique, given adequate mixing and convergence, has the advantage of directly sampling from $p(\mathcal{M}|d)$
as opposed to approximating or calculating the model evidence.  Like its fixed dimension counterpart, the RJMCMC is
guaranteed to converge to the appropriate PDF.  The model posterior is determined by the number of iterations the chain
spends in each model (normalized by the total number of iterations).  With a small number of models present (as in our example)
a more useful representation of the information garnered by the RJ algorithm is to calculate the Bayes factor for two competing models
\begin{equation}
B_{10}=\frac{\text{\# of iterations in }\mathcal{M}_1}{\text{\# of iterations in }\mathcal{M}_0}.
\end{equation}
Allowing for the exploration of different models (which may differ in dimension) requires a separate Hastings step which
proposes to move the chain from one model to another.  Parameters for trial model $\mathcal{M}_i$ are drawn from
$q(\vec{\theta}_i, \mathcal{M}_i)$.  If the models are nested (perhaps proposing additional parameters to include in
the existing set) all of the like parameters are held fixed while the new parameters are drawn from $q$.
In the detection problem case, where the models are disjoint, all model parameters must be randomly selected.
Once the new model's parameters are in hand the trans-model Hastings ratio (again, satisfying detailed balance) is calculated by
\begin{equation}
H_{\mathcal{M}_i \rightarrow \mathcal{M}_j}=\frac{p(s|\vec{\theta}_j, \mathcal{M}_j)p(\vec{\theta}_{j},  \mathcal{M}_{j})q(\vec{\theta}_{i},\mathcal{M}_{i})}
	    {p(s|\vec{\theta}_i, \mathcal{M}_i)p(\vec{\theta}_{i},  \mathcal{M}_{i})q(\vec{\theta}_{j},\mathcal{M}_{j})} |J_{ij}|
\end{equation}
where the Jacobian $|J_{ij}|$ accounts for any change in dimension between models $\mathcal{M}_i$ and $\mathcal{M}_j$
which are parameterized by $\vec{\theta}_i$ and $\vec{\theta}_j$ respectively.  Selecting an efficient proposal distribution
for model transitions is typically the major obstacle in the implementation of an efficient RJ routine.  If the proposal
distributions yield the model parameters directly, instead of a set of random numbers which are then used to determine
the new model parameters, the Jacobian is unity and can be neglected.  RJMCMC algorithms are notoriously difficult to
implement because dimension changing moves are typically accepted with very low frequency causing the chains to
converge slowly.  This problem is exacerbated when competing models are of high dimensionality.  It is therefore vital
to the success of the algorithm to construct trans-model proposals which allow for decent mixing between the two models.
Without adequate between-model transitions the inferred Bayes factors are suspect.

\subsubsection{Transdimensional Proposal Distribution} \label{hypercube}
It took us many attempts to get the RJMCMC routine to work. Our initial efforts used the Fisher approximation to the posterior
in the neighborhood of the MAP parameters found from the PTMCMC search/characterization.  This technique showed initial
promise by producing qualitatively reasonable Bayes factors (monotonically increasing as a function of SNR) and stability
(producing similar results through different random seeds).  However, further testing revealed disagreement with the
thermodynamic integration and erratic behavior including large increases in Bayes factor for very small increases in SNR,
and occasionally questionable results (such as ``detectable''  sources with SNR well below five).  This was confounding
until we noticed that the RJMCMC chains were producing PDFs that differed significantly from the PTMCMC results. This
signaled poor exploration of the pror and a lack of convergence towards the true distribution.  It wasn't until
the Fisher approximation became sufficiently accurate (the SNR became sufficiently large) that this proposal distribution
mimicked the target distribution well enough to allow for adequate mixing.  It was difficult to immediately diagnose, based on the behavior
of a RJ chain alone, what the problem was, and without having the thermodynamic integration results for comparison,
it would have been difficult to know if alternative approaches fixed the problem.  

We settled on a method for constructing  trans-model proposals first suggested by Green \cite{Green2003} to start with fixed dimension MCMC
trial explorations of each model in question.  The resultant joint PDF from this learning period can be used as the proposal
distribution used to move into that model. The implementation of this scheme which would yield the most efficient between-model
exploration would be to sample from the fixed-dimension joint PDF directly.  Unfortunately, thoroughly sampling the PDF for
high dimension models rapidly becomes prohibitively costly.  Instead, we approximate the full PDF using the chain from the
PTMCMC analysis used in the initial detection.  

For noise parameters this is simply a normal distribution for each frequency window and channel with mean
$\eta^i_{\alpha,\text{MAP}}$ and variance $1/N$.  The mean of the noise-parameter proposal distribution is the
maximum \emph{a posteriori} value for each parameter found from the fixed-dimension PTMCMC.  This approach requires
$\mathcal{M}_1$ and $\mathcal{M}_0$ to have slightly different trans-model proposal distributions for noise parameters
as the analysis done under different models will yield different MAP noise parameters.

Sufficiently sampling the full signal posterior would require ~$10^{10}-10^{12}$ samples, whereas our typical Markov
chains are only of length ~$10^6$ samples.   Clearly the full PDF will be woefully under-sampled, but thanks to the
excellent mixing from the fixed dimension exploration (courtesy of the parallel tempering) we are confident that the
high probability density regions of the PDF \emph{have} been adequately sampled and qualitatively reflect the true
joint distribution.  To construct the signal parameter proposal distribution $q(\vec{\lambda},\mathcal{M}_1)$ we first
divide the prior volume for each parameter $\lambda^i$ by the average standard deviation $\bar{\sigma}^i_{\Gamma}$ of
that parameter.  This average is taken over the parameter variances calculated each time the Fisher Information Matrix
is updated during the PTMCMC analysis. 

The prior volume can now be thought of as (for the galactic binary parameterization) an 8-dimensional hypercube with
cells of width $\bar{\sigma}^i_{\Gamma}$.  The volume of each cell is  $dV = \prod_i \bar{\sigma}^i_{\Gamma}$ and the
total number of cells in the hypercube is $N_{\text{cell}}=\prod_i \Delta p(\lambda^i)/\bar{\sigma}^i_{\Gamma}$ where
$\Delta p(\lambda^i)$ is the prior range for parameter $i$.  We then sort through the fixed dimensional signal-model
Markov chain from the PTMCMC to populate the cells in the hypercube.  Typically between $10^3$ to $10^5$ cells are
actually occupied for fixed-dimension chains of length $N_{\text{MCMC}}=10^6$.  If $n$ points in the chain land in
the cell with reference parameters $\vec{\lambda}$ then the probability density of jumping into that cell is
\begin{equation}
q_s(\vec{\lambda},\mathcal{M}_1) = \frac{n(\vec{\lambda})}{dVN_{\text{MCMC}}}
\end{equation}
This is not yet a valid proposal distribution, however, because drawing from $q_s(\vec{\lambda},\mathcal{M}_1)$ will not
allow the signal model transitions to access the entire prior volume.  To rectify this we add to the existing distribution
a uniform distribution over the entire cube, by giving each cell a single occupant.  The uniform piece of the proposal
has probability density
\begin{equation}
q_u(\mathcal{M}_1)=\frac{1}{dV N_{\text{cell}}}
\end{equation}
The full proposal distribution is then the sum of the sampled and uniform cell populations, with relative weighting $w$ 
\begin{equation}\label{trans-D}
q(\vec{\lambda},\mathcal{M}_1) = w q_s + (1-w)q_u.
\end{equation}
To ensure proper normalization of the proposal distribution $w \in [0,1]$. 

To draw from this hybrid distribution we first compare a number from $U[0,1]$ to $w$.  If this exceeds $w$ we uniformly
draw the signal parameters from the prior.  We then must determine to which cell of the hypercube the new set of
parameters belongs so we may evaluate $q(\vec{\lambda},\mathcal{M}_1)$.  When drawing from the PTMCMC-populated
portion of the distribution we randomly select one ``link'' from the fixed dimension chain to select an occupied cell
of the hypercube.  This guarantees that moves to a particular occupied cell are proposed at the same frequency with
which the cell was sampled during the search.  We then draw the signal parameters uniformly from within the selected
cell.  For transitions into $\mathcal{M}_0$ we must again determine where in the cube the current signal parameters
reside to accurately calculate the trans-dimensional Hastings Ratio.

\subsubsection{RJ Parallel Tempering}
Between trans-dimensional proposals the chains update within their current dimension identically to the description described
in \S\ref{mcmc}.  We only allow the $\beta=1$ chain to undergo trans-dimensional moves.  When this chain is in a given model
it has $N_c \sim 20$ chains of the same model, each of increasing temperature, with which to exchange parameters.  When
trans-dimensional moves are accepted by the ``cold'' chain these parallel tempered chains stop updating and store their
current locations in parameter space, while the analogous chains in the other model resume from their most recent positions.
This way the chain sampling from the target distribution is always in contact with a ``heat reservoir'' of parallel chains,
but we can avoid wasting computations for chains which have no way of communicating with the $\beta=1$ parameters.  The
schedule for these different types of updates (inter-dimensional MCMC, parallel tempering parameter exchanges, and
trans-dimensional proposals) is part of the overall proposal distribution and therefore does not affect the outcome.
This schedule does, however, need to be tuned in order to see convergence within a reasonable amount of time.

\section{Results} \label{results}
Three test sources, chosen to provide representaitve examples over the sky and frequency range of compact binaries, were
constructed and injected into several different noise realizations per signal.  

\begin{table}[htdp]
\caption{Injected galactic binary parameters. }
\begin{center}
\begin{tabular}{c c c c c c c c  }
	\toprule
	Source & $f$ (mHz) & $\cos{\theta}$ & $\phi$ & $\cos{\iota}$ & $\psi$ & $\varphi_0$ & $\dot{f}T^2_{\text{obs}}$ \\
	1 & 1 & 0.713 & 0.452 & 0.534 & 2.334 & 0.624 & 3.241 \\
	2 & 3 & 0.598 & 2.972 & 0.827 & 1.240 & 5.938 & 0.139 \\
	3 & 5 & 0.326 & 4.644 & 0.133 & 0.314 & 3.878 & 0.643 \\
\end{tabular}
\end{center}
\label{params}
\end{table}%

For each case the analysis was repeated for a range of amplitudes, corresponding to signals with a range of signal-to-noise ratios
defined by
\begin{equation} \label{snr}
\text{SNR} \equiv \frac{(s|h(\vec{\lambda}))}{\sqrt{(h(\vec{\lambda})|h(\vec{\lambda})}}
\end{equation}
We are then able to learn how the Bayes factor increases as a function of SNR to infer when that particular source, in
that particular noise realization, becomes ``detectable'' ($B_{10} \gtrsim 3$).  The data was analyzed using several runs of
the fixed dimension PTMCMC (one for each model) with different seeds used to initiate the chains, thus testing the stability
of the evidence calculation.  We used 20 chains in the parallel tempering scheme with a maximum temperature of 100.
The temperature spacing of the chains was geometric in $\beta$.  

\subsection{Search Phase}
\begin{figure}[htbp]
   \includegraphics[height=0.5\textwidth, angle=270]{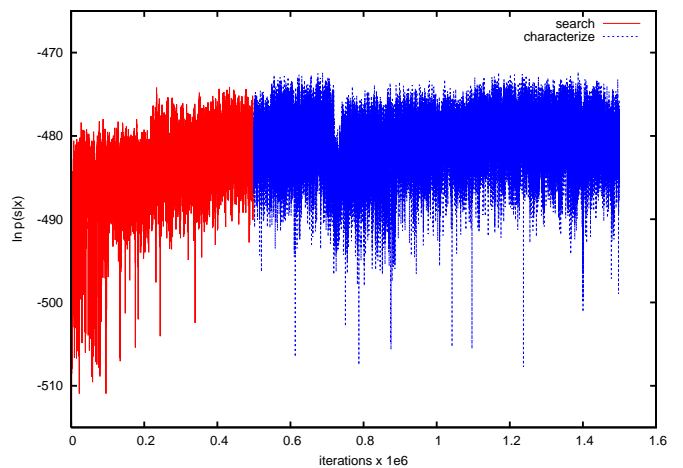} 
   \caption{$\ln p(s|\vec{\theta})$ during search (red, solid) and characterization (blue, dashed) phase of
the analysis for a SNR $=8$ source.}
   \label{search}
\end{figure}

The chains were initialized randomly within the parameter space and allowed $5 \times 10^5$ iterations of ``burn-in''
before the samples were considered to be accurately representing the target distribution.  As can be seen from an
example chain in Figure~\ref{search} the $\beta = 1$ chain reached a stationary solution in the vicinity of the
maximum log-likelihood within the burn-in phase (the red [solid] portion of the plot) thus signaling the arrival
of the search to the region(s) of maximum posterior weight.  This demonstrates the success of the search phase
in a reasonable number of iterations without non-Markovian convergence aids.

\subsection{Characterization}
\begin{figure*}[htbp]
   \centering
   \includegraphics[width=\textwidth]{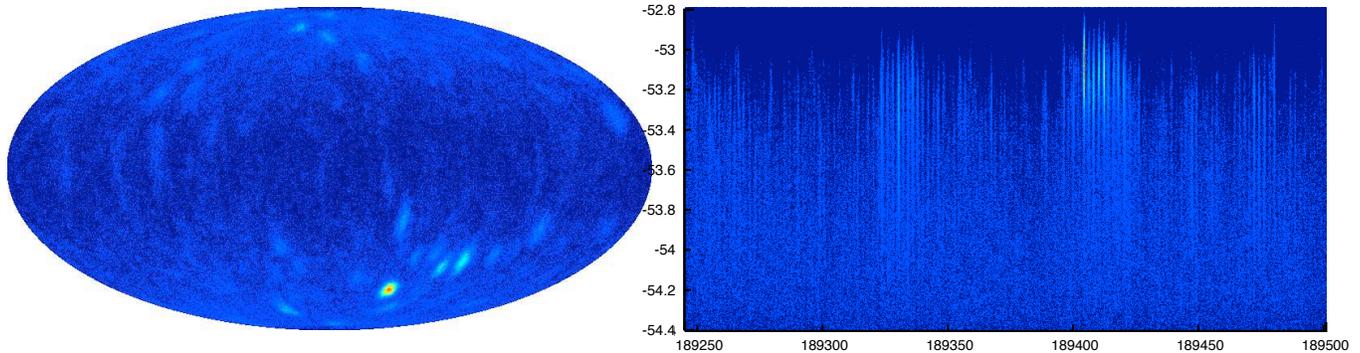} 
   \caption{Marginalized log posterior densities in the $\theta-\phi$ (left) and the $\ln A-fT_{\text{obs}}$ (right) plane for data containing only stationary-gaussian noise.  Although this data contained no gravitational wave signal the PDFs show organized locations in parameter space that \emph{look} like GW signals.  The evaluation step of the analysis easily discarded these potential detections returning a Bayes factor of $\sim 0.7$.  The $\theta-\phi$ PDF is shown in a Molweide projection on the sky.  }
   \label{snr0}
\end{figure*}

The $10^6$ iterations post burn-in were used to characterize the candidate detection.  The parallel tempering
allowed the chains to thoroughly explore the entire posterior distribution function within this alloted time.  We were
surprised to see a great deal of structure in the signal posterior even in data where no signals were injected. 
Figure~\ref{snr0} shows the marginalized posterior distributions for both the sky location and the $fT-\ln A$ plane when
the data contained only simulated noise.  Notice the concentration of posterior weight at one localized sky location and frequency bin,
despite the data being devoid of any injected signal.  The MAP parameters recovered by the analysis were for a (false positive) source with $\rm{SNR}\sim6$ while the evidence ratio between this and the noise-only model was $\sim 0.7$.  This strengthens our opinion that rigorous model selection steps are
necessary, as even the analysis of noise-only data recovered a ``believable'' signal which was well localized in parameter
space but easily rejected by our Bayes factor calculations.  PDFs of higher signal-to-noise sources continue to demonstrate
complicated secondary features in addition to the regions of parameter space near to the injected parameters.
Figure~\ref{ptsky} (bottom panel) in \S~\ref{parallel tempering} is an example of this multi-modality.

\subsection{Evaluation}
\begin{figure}[htbp]
   \centering
   \includegraphics[height=0.5\textwidth, angle=270]{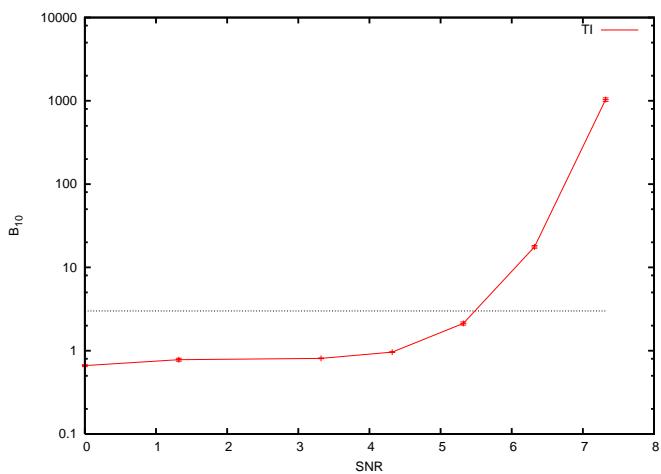} 
   \caption{Thermodynamic integration results for $B_{10}$ on data with signals injected at increasingly higher signal-to-noise ratio.  Error bars were established by starting the chains with different random number generator seeds.  The horizontal line is the Bayes factor where one would consider the result a positive detection. }
   \label{ptbxy}
\end{figure}

\begin{figure}[htbp] 
   \centering
   \includegraphics[height=0.5\textwidth, angle=270]{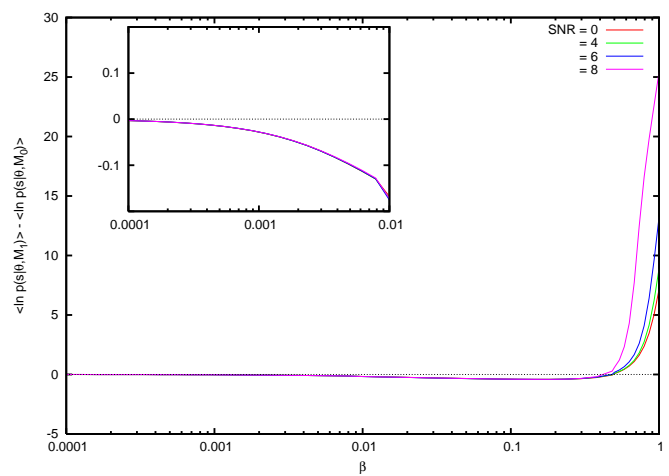} 
   \caption{Integrand for thermodynamic integration on data with signals injected at increasingly higher SNR.  Of importance is how the curves become indistinguishable below $\beta \sim 0.01$ [inset].  The horizontal line marks the regimes where the integrand supports $\mathcal{M}_1$ (above) or $\mathcal{M}_0$ (below).  The inset shows that the integrand has sufficiently vanished at the maximum temperature analyzed.}
	\label{thermo_convergence}
\end{figure}

The model evidence for each SNR and each hypothesis ($\mathcal{M}_0$ and $\mathcal{M}_1$) is calculated using the 20 chains
from the fixed dimension exploration with the integral over $\beta$ being evaluated from $10^{-2}$ to $1$.
Figure~\ref{ptbxy} shows, for a single source and noise realization, the Bayes factor as a function of increasing SNR.
The source becomes ``detectable'' ($B_{10}\gtrsim3$) in the vicinity of the fiducial signal-to-noise ratio of five threshold.
Figure~\ref{thermo_convergence} shows the integrand from eq.~\ref{thermo_bayes_integral} for injected signals of
SNR~$=[0,8]$ (The signals were injected with amplitudes set using the definition SNR~$ \equiv \sqrt{(h|h)}$ which differs
from Eq. \ref{snr}.  It is expected that the SNRs when calculated via Eq. \ref{snr} and $\sqrt{(h|h)}$ will differ
(for data comprised of a finite number of frequency bins).  The results of the Bayes factor vs. SNR studies use
Eq. \ref{snr} to ensure that the interpretation of the SNR is the same between noise realizations.)  

From this figure we can learn much about thermodynamic integration, including how a maximum temperature of 100 is
clearly insufficient to capture all of the weight in the integrand.  Running chains out to a
more appropriate temperature ($T \sim 1 \times 10^4$) with sufficiently dense spacing is inconvenient.
This problem can be circumvented, however, because at sufficiently high temperature the expectation value of the
likelihood becomes independent of the injected signal.  This manifests itself in the Figure~\ref{thermo_convergence}
as the different curves become identical at lower $\beta$.  This is easily understood if one considers that the
$1/T$ term in the likelihood decreases the effective SNR of the injected source by $\sim 1/\sqrt{T}$.  Because our
analysis did not extend beyond sources with SNR $= 10$, a maximum temperature of 100 ($\beta = 0.01$) sufficiently
hides any contribution to the data by the gravitational wave.  Therefore we can perform the evidence integration
up to $\beta=0.01$ once for each noise realization and only need to do the case by case study in the low temperature
regime.  To take advantage of this the analysis was repeated once for each noise realization with 20 chains spaced
geometrically with [$T_{\text{min}},T_{\text{max}}] = [100,10000]$.  The evidence ratio from this high temperature analysis
is a common multiple which needs to be applied to the Bayes factors found for each SNR of that noise realization.  The inset in Figure~\ref{thermo_convergence}
shows that the integrand has very nearly gone to zero at the maximum temperature of our analysis, signaling convergence of
the integral.

\subsection{Cross Validation}

\begin{figure}[htbp]
   \centering
   \includegraphics[height=0.5\textwidth, angle=270]{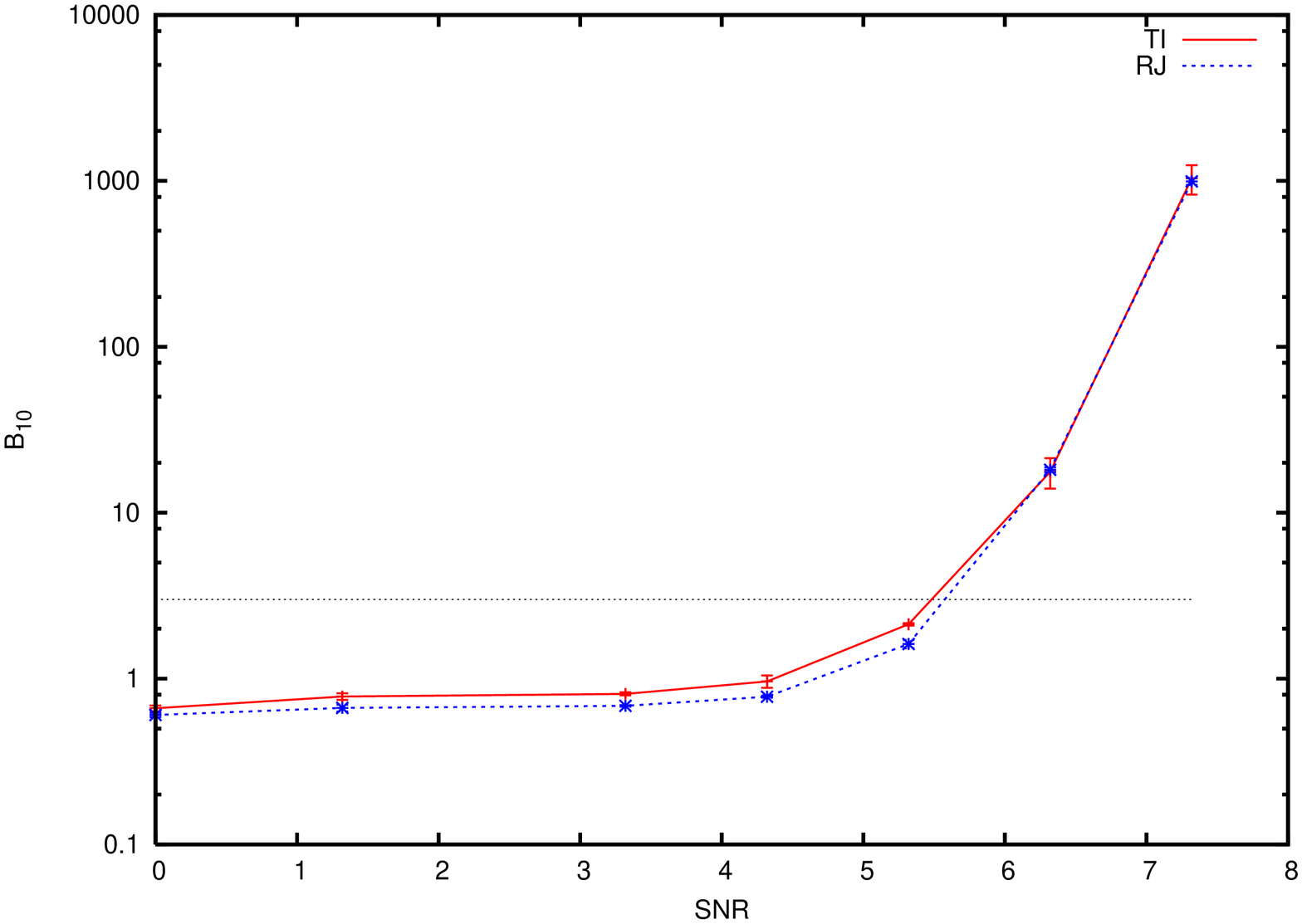} 
   \caption{Thermodynamic integration and RJMCMC results for $B_{10}$ on data with signals injected at increasingly higher signal-to-noise ratio.  The horizontal line is the Bayes factor where one would consider the result a positive detection.  }
   \label{rjbxy}
\end{figure}

To verify the Bayes factor calculation from the thermodynamic integration the post burn-in samples from the $T=1$ chain
are sorted into the signal-model hypercube that will be used for the trans-dimensional proposal distribution in the RJMCMC.
We experimented with a variety of different weightings for the sampled and uniform portions of the proposal, and found little
variation in the resultant Bayes factors.  Because the result is supposed to be independent of the proposal distribution
this serves as a cheap test to ensure stability and convergence.  For production-level runs the relative weighting between
the two contributions to the proposal distribution was $w=0.5$ although different values of $w$ were used in testing to help verify that the proposal distribution was not biasing the results.  The first $10^5$ RJ iterations were burn in samples,
after which the samples were assumed to be drawn from the model posterior.  Figure \ref{rjbxy} shows the achieved agreement
between the TI and RJMCMC calculations of the Bayes factor as a function of increasing SNR.  

\subsection{Dependence on Priors}
\begin{figure}[htbp]
   \centering
   \includegraphics[height=0.5\textwidth, angle=270]{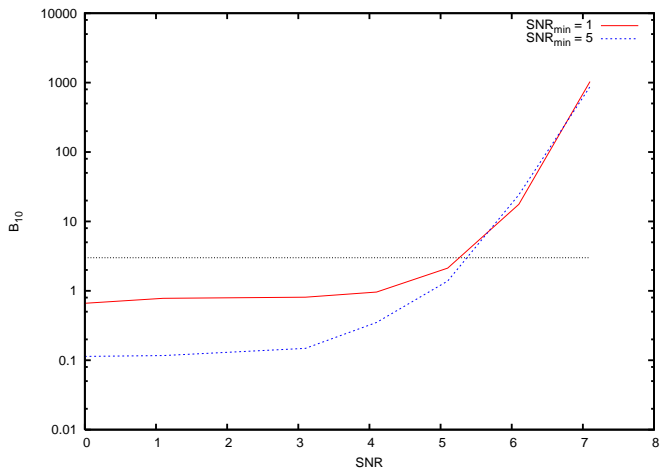} 
   \caption{Thermodynamic integration results for $B_{10}$ on data with signals injected at increasingly higher signal-to-noise ratio.  The horizontal line is the Bayes factor where one would consider the result a positive detection.  The red (solid) line is for the analysis done with uniform priors in the signal parameters (apart form $\dot{f}$).  The blue (dashed) line is the same data analyzed with a restrictive prior on amplitude such that the minimum Amplitude corresponds to a SNR of $\sim 5$.  See \S\ref{prior} for more explanation.}
   \label{r_prior}
\end{figure}

As discussed in \S\ref{prior} the analysis was repeated adopting an amplitude threshold such that the minimum allowed
template SNR was $\sim 5$.  Although the SNR at which the signal became detectable was unchanged, the restrictive prior volume
did allow the analysis to definitively prefer the noise model over the signal model when the injected signal was well
below this threshold.  This is in opposition to the examples which allowed an arbitrarily small amplitude, which caused
the two models to be indistinguishable at very low SNR despite the additional degrees of freedom of the signal model.
Figure \ref{r_prior} shows the Bayes factor as a function of SNR for the two cases with the red (solid) line depicting
the results with the uninformative prior while the blue (dashed) line resulted from the constrained amplitude prior case.
What should be taken from this demonstration is how the results of Bayesian methods depend upon what is being asked of the
data.  The two questions asked here, ``Does the data contain \emph{any} gravitational waves?'' and ``Does the data contain
gravitational waves \emph{above a certain amplitude}?'' have very different interpretations and thus result in very
different responses from the analysis.  This is a feature (rather than a flaw) of Bayesian methods that underlines the role and importance of ones \emph{a priori} beliefs about the experiment.

\section{Discussion} \label{discussion}
We have implemented and tested an end-to-end Bayesian analysis algorithm which proved successful at solving the detection problem
for a single galactic binary signal in simulated LISA data with stationary, gaussian noise.  We see our approach as a first step towards the ultimate goal of generically solving the detection problem in a fully Bayesian manner for realistic data (i.e., non-stationary, non-gaussian noise and source confusion).  

Our algorithm single-handedly locates the regions in parameter space of
high posterior weight (search phase), thoroughly samples the distribution about these locations (characterization phase),
and quantifies the confidence of the detection (evaluation phase).  All three phases of the detection problem have been
successfully completed using a single routine which resolves the PDF and calculates the evidence for each of the models
under consideration.  The construction of such an approach is relatively simple, and this simplicity will serve as the foundation for a
reliable ``one stop'' solution to the detection problem.     

The search and characterization phases were simultaneously accomplished through a parallel tempering MCMC analysis
which located the regions of interest within a few hundred thousand iterations.  The efficient exploration of the
full posterior was clearly demonstrated as the results revealed a surprising amount of additional structure in the signal PDFs.
This could conspire to ``fool'' a less sophisticated analysis by preferentially sampling from some secondary maxima,
thus failing to properly characterize the source.  This could lead to a poor estimation of astrophysical parameters,
or worse, a missed or false detection.

Our algorithm was \emph{not} fooled thanks to the model selection analysis provided by the thermodynamic integration across
the PTMCMC chains. The performance of the model selection phase of the problem was verified with a RJMCMC
analysis. It is our belief that thermodynamic integration has proven to be the most reliable technique, and is exceptionally
appealing when we consider extending this algorithm to a more general GW detection problem.  The implementation of the
thermodynamic integration calculation is independent of the model dimension and should remain as robust for more
complicated scenarios.    

Although we also had success with the RJMCMC algorithm, we caution that the RJMCMC mixing and convergence time was very sensitive
to the construction of the algorithm, and is prone to ``fooling'' the user by appearing to converge to a stationary distribution
that is \emph{not} the target distribution.  The diffuculty, a well documented one in the employment of the RJMCMC, is constructing
an efficient trans-dimensional proposal distribution.  Although the proposal is not supposed to affect the end result of the
analysis, we found that sub-optimally tuned proposals led to very poor inter-dimensional mixing which in turn leads to very
slow convergence:  Well beyond the $10^6$ iterations used to generate the displayed results.  Once we adopted the PTMCMC-PDF proposal
it was discovered through repeated testing that the chains were stable and rapid in their convergence and in good agreement
with thermodynamic integration.  Figure~\ref{rjbxy} demonstrates the RJMCMC's robustness against different starting seeds
under this efficient trans-dimensional proposal scheme.

Proving that we are recovering an unbiased set of samples from an unknown target distributions is a subtle, if not impossible, task.  There is a standard suite of convergence tests to asses the stationarity of the chains, and the degree of correlation between chain samples.  However, chains which are strictly sampling from a single mode of an otherwise multi-modal distribution can pass these tests making it difficult to quantify the global accuracy of the recovered distributions.  

\begin{figure*}[htbp]
   \begin{center}
   \includegraphics[width=\textwidth]{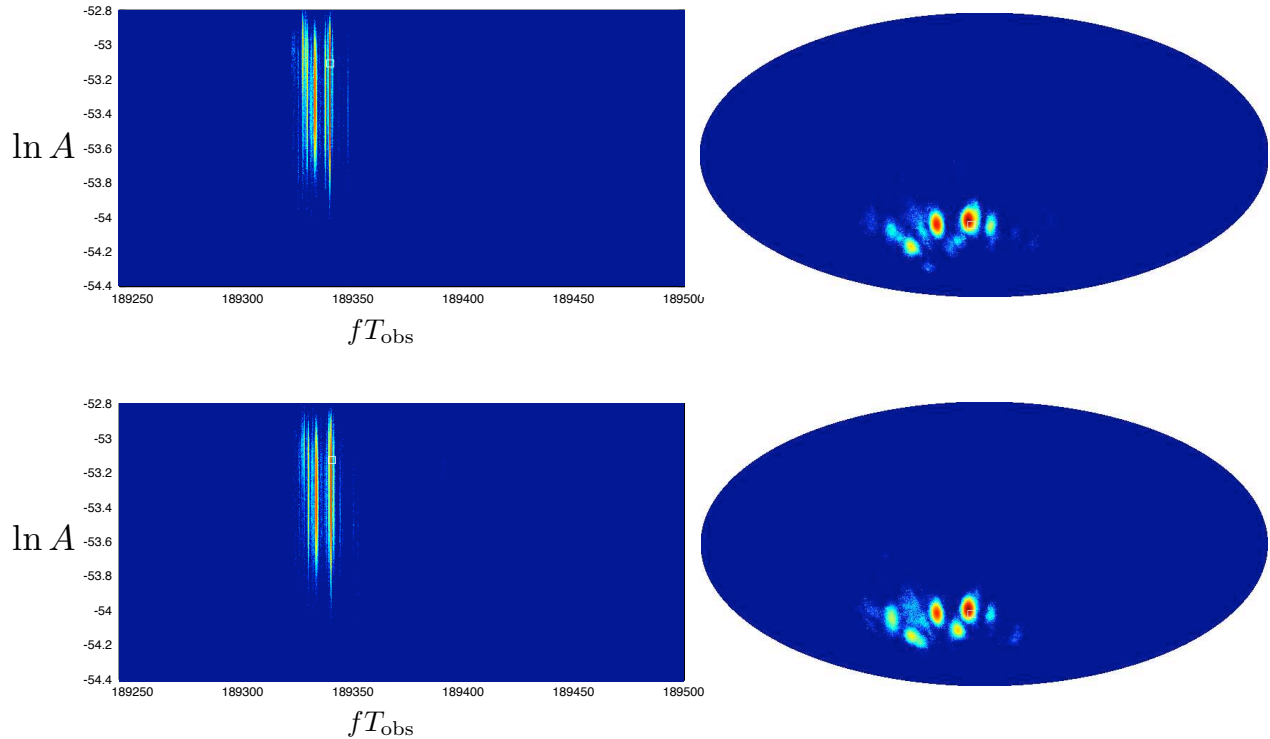} 
   \caption{Marginalized log posterior densities for sky location (left) and $fT_{\rm{obs}}-\ln{A}$ (right) for two runs
using different random number seeds for the Markov chains.  The consistency between different runs is evidence that the chains have
converged to target distribution.}
   \label{convergence}
  \end{center}
\end{figure*}

To test for convergence of the PTMCMC and RJMCMC chains we use variants of the Geweke convergence test (checking that early and
late sub-samples of the chains produce the same distributions) and the
Gelman and Rubin approach (comparing many repeated runs using different
random seeds - and hence widely dispersed starting locations) as shown in Figure~\ref{convergence}. The
Geweke test is a simple approach to seeing if a particular chain has reached
stationarity, but as stated previously, chains that are trapped at local modes can pass this
test.  To asses convergence of the global structure of the posterior
distribution we demand that the recovered PDFs are qualitatively similar between multiple runs with widely seperated starting
points.

The comparison of Bayes factor results between TI and RJMCMC is not a fully independent consistency check, as the RJMCMC
empoyed the fixed dimension PDFs derived from the PTMCMC runs as part of the trans-dimensional proposal distribution.
While, in theory, the MCMC algrotihm will return the correct posterior distribution independent of the proposal
distribution, in practice, for any finite length run one may end up with a distribution that has been biased by the
choice of proposal distribution. This bias was apparent when we experimented with trans-dimensional proposals based on
Gassian (Fisher Matrix) approximation to the fixed dimension PDFs.

To explore the dependence on the choice of trans-dimensional proposal distributions we performed multiple runs using
different weightings for the two contributions to the trans-dimensional proposal distribution described in~\S\ref{hypercube}.
The RJMCMC results were found to be consistent for values of $w$ in Eq.~\ref{trans-D} between 0.3 and 0.9. Taking
$w$ below 0.3 gave very poor acceptance rates and hence long convergence times, so we were not able to check for
consistency at very low values of $w$. We also did not use $w$ values above 0.9 as some fraction of the distribution has to
cover the full prior volume. While these tests do not prove that our RJMCMC results are fully independent of the
PTMCMC results, the results show no obvious dependence on the choice of proposal distribution.  Furthermore, it is
worth emphasizing that the fixed dimension proposal distribution was only used for trans-dimensional moves.  Between
trans-dimensional moves the chains were free to explore the parameter-space of their respective models using the full
cocktail of fixed-dimension proposal distributions.  Among these fixed-dimension proposals are large jumps (such as uniform
draws on the full prior range) which are included to facilitate the chain's global exploration. The inclusion of
parallel tempering also helps to ensures that the chains are not ``trapped'' in regions favored by the trans-dimensional
proposal distribution.

To test further the model selection phase of the problem we experimented (unsuccessfully) with constructing a
Nested Sampling \cite{Skilling1, Skilling2} evidence calculator.  Nested Sampling is receiving attention in the astrophysics and
cosmology community (e.g. \cite{Mukherjee, Feroz}) and more recently in gravitational wave detection problems similar to the
one considered here~\cite{Veitch1, Veitch2}.  Nested Sampling is an alternative to the MCMC-driven methods discussed previously.
The algorithm stochastically explores the posterior by first drawing from the prior distribution with some fixed number of
samples.  The ``worst-fit'' sample is replaced by one with higher likelihood (found using a hill climbing routine or even
a MCMC algorithm), while also reducing the volume covered by the samples. As this procedure is iterated the samples become
tightly clustered in the regions(s) of high posterior weight.  The evaluation of the evidence reduces eq.~\ref{evidence integral}
in \S\ref{bayes} to a one-dimensional integral over the fraction of the prior volume contained within the iteratively more
restrictive likelihood constraint.  A more detailed description of the step-by-step workings of the algorithm can be found
in Ref. \cite{Skilling2}.

A straighforward implementation of the Nested Sampling algorithm gave results that varied greatly from run to run. In
many instances one or more of the maxima of the posterior would be missed entirely, and different methods for estimating
the fraction of the prior volume covered by the samples gave very different results. In an effort to improve the
performance we used  the PDFs derived from the PTMCMC algorithm to determine which regions of parameter space need to be included
in the evidence calculation. Regions of parameter space not included in the signal-model hypercube (see \S\ref{hypercube})
were assumed to give a negligible contribution to the evidence integral. The Nested Sampling algorithm was applied to
each occupied cell, and the evidence
in each cell was added together to give the total evidence for the model. While this trick ensured that the regions of
high posterior weight were included in the integration, the results were still highly dependent on how we calculated the
fractional volume covered by the samples.

We have recently learned of a new Nested Sampling algorithm called MultiNest~\cite{Feroz:2007kg, Feroz:2008xx} which provides a
solution to the problem of computing the volume covered by the samples and a reliable technique for finding new samples
with higher likelihoods that can handle multi-modal posteriors. The MultiNest algorithm has been sucessfully applied to
high-dimension parameter estimation problems in gravitational wave data analysis~\cite{Gair:2009cx, Feroz:2009de}. 
Even with this improved algorithm, it is necessary to first identify the rough location of the posterior modes in order to
keep the computational cost down. In future work we hope to compare the evidence estimates from the MultiNest algorithm to
those from TI and RJMCMC.

When comparing the two model selection techniques that we have successfully implemented (TI and RJMCMC), we found TI
to be the simplest to implement and most robust in terms of convergence for a wide range of proposal distributions. In
our experience the RJMCMC technique required very carefully designed proposal distributions to achive reasonable
acceptance rates for the trans-dimensinal moves, and the results can be biased by poor choices for the trans-dimensinonal
proposals.

\begin{figure}[htbp]
   \centering
   \includegraphics[height=0.5\textwidth, angle=270]{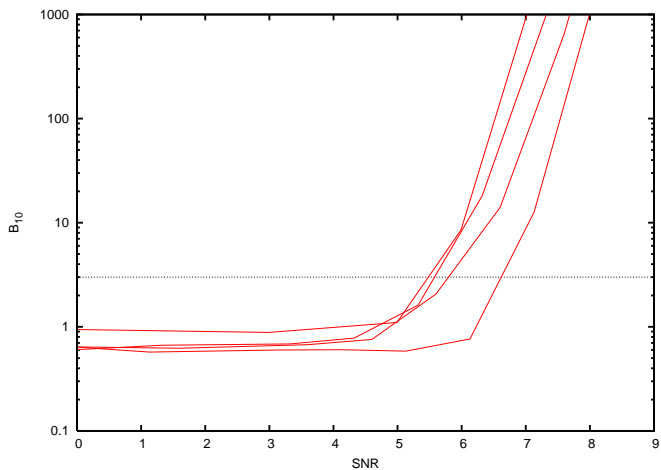} 
   \caption{Thermodynamic integration results for $B_{10}$ on data with signals injected at increasingly higher signal-to-noise ratio.  The horizontal line is the Bayes factor where one would consider the result a positive detection.  Different curves represent different noise realizations but identical signal parameters.  This demonstrates that the detection ``threshold" is sensitive to the particulars of the noise in the detector, even when the noise characteristics are identical (stationary, gaussian, with known spectral density).}
\label{bxy_noise}
\end{figure}

\begin{figure}[t]
   \centering
   \includegraphics[height=0.5\textwidth, angle=270]{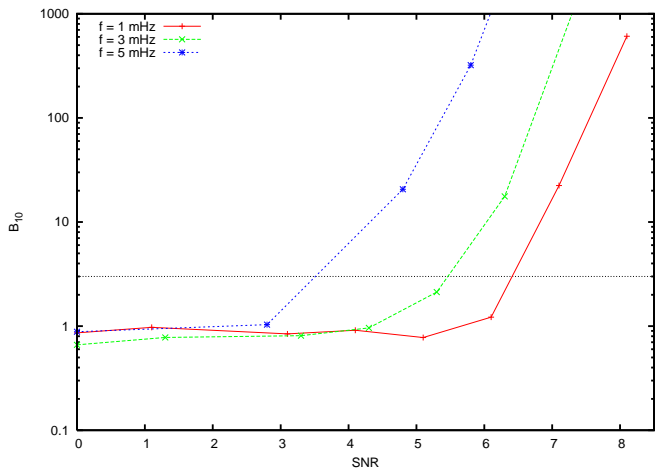} 
   \caption{Thermodynamic integration and RJMCMC results for $B_{10}$ on data with injected signals 2, and 3 (red \& green and blue \& magenta, respectively) injected at increasingly higher signal-to-noise ratio (see Table \ref{params}.  The horizontal line is the Bayes factor where one would consider the result a positive detection.  This demonstrates that the detection ``threshold" is sensitive to the details of the waveforms.}
\label{bxy_source}
\end{figure}

By comparing the SNR at which a signal becomes detectable between different noise realizations the weakness of using
fixed SNR detection thresholds becomes apparent.  Figure~\ref{bxy_noise} shows the Bayes factor as a function of
injected SNR for three identical sources injected into three different noise realizations.  The point at which a
signal becomes detectable varies appreciably between noise realizations even for benign, stationary and Gaussian noise.
Even under ideal circumstances the detectability of a source is dependent on the particular noise realization
with which the source is competing, as opposed to a hypothetical ensemble of possible noise realizations.
Predictably, similar variation in source ``detection thresholds''  comes from the details of the waveforms themselves, as can
be seen in the Bayes factor vs. SNR plots of the different test signals (fig. \ref{bxy_source}).

In this study we have ignored many real world complications.  Absent a good noise model for LIGO-Virgo or LISA the results taken from this algorithm must be treated with caution. Like any calculational
tool, the answers received are only as good as the models under comparison.  The Bayes factors that we compute do not answer the question ``is there a GW signal present in the data".  Both models are imperfect because our description of the LISA instrument noise is grossly over simplified.  Addressing the question we really want to answer - ``is there
a binary white dwarf signal in this stretch of data" - will require a more realistic noise model.  That being said, we feel that this work is a useful contribution towards providing a solution to the detection problem, particularly in calculating the evidence for large-dimension models.

In addition to our simplified noise modeling, a 256-bin segment of LISA data will
likely contain several galactic binaries with varying degrees of overlap.  To properly address the LISA-specific detection
problem we would need to allow for an arbitrary number of sources and our algorithm would be responsible for sampling the
model posterior $p(\mathcal{M}_i, s)$ to determine the most likely number of sources, instead of the binary 0- or 1-source case.
An example of this type of study made with sinusoidal waveforms in simulated LISA data can be found in Ref.~\cite{Umstatter}.
The RJMCMC approach nicely generalizes to the ``$N$ or $N+1$ sources'' detection problem.  One problem that has been faced
while doing global galaxy fitting is the occasional tendency for the algorithm to fit a single galactic binary with
two highly-overlapping waveforms. There are specific trans-dimensional moves, split/merge proposals \cite{Umstatter}, which are designed to resolve problems with overlapping signals that can be applied the RJMCMC algorithms. The model evidence, with its built in Occam factor, will favor solutions where a single
waveform template is used to model each signal.  In future work we will extend our modeling to include multiple, overlapping signals

Our single-source toy problem is more akin to the ground-based application of model selection where GW events are expected to be rare.
However, our toy problem is unrealistic because our treatment of the noise does not allow for non-stationarity or departures from
Gaussianity.  

One possibility would be to treat the evidence as a detection
statistic that can be computed from the data and calibrated using signal
injections and time slides.  Instead, we would prefer to ``calibrate" the evidence
by more accurately modeling the noise. There are a variety of ways in which this could
be done, and we are currently pursing several options, such as
comparing the evidence between different noise models, or using parameterized
noise models (that allow for tails, etc.) and finding the parameters that
best fit the noise. While there is no way of knowing if the ``correct" noise
model has been found, the analysis in Ref.~\cite{Allen} suggests that a fairly
wide class of noise models should give adequately promising performance.

\section{Acknowledgments}
This work was supported by NASA grant NNX07AJ61G. We thank Kipp Cannon and Antony Searle for
providing helpful comments on an earlier draft of the paper.

\end{document}